\documentclass[twocolumn,prl,showpacs,preprintnumbers,amsmath,amssymb]{revtex4}

\usepackage{graphicx}% Include figure files
\usepackage{dcolumn}% Align table columns on decimal point
\usepackage{bm}% bold math
\usepackage{subfigure}
\usepackage{float}

\begin{document}

%\preprint{APS/123-QED}
\title{\textbf{\textrm{Structural versus dynamical origins of mean-field behavior in a self-organized critical model of neuronal avalanches}}}

\author{S. Amin Moosavi}
\author{Afshin Montakhab}
    \email{montakhab@shirazu.ac.ir}

\affiliation{Department of Physics, College of Sciences, Shiraz University, Shiraz 71946-84795, Iran}

\date{\today}
%%%-----------------------------------------------------------------------------
\begin{abstract}
Critical dynamics of cortical neurons have been intensively studied over the past decade. Neuronal avalanches provide the main experimental as well as theoretical tools to consider criticality in such systems. Experimental studies show that critical neuronal avalanches show mean-field behavior. There are \textit{structural} as well as recently proposed [Phys. Rev. E \textbf{89}, 052139 (2014)] \textit{dynamical} mechanisms which can lead to mean-field behavior. In this work we consider a simple model of neuronal dynamics based on threshold self-organized critical models with synaptic noise. We investigate the role of high average connectivity, random long range connections, as well as synaptic noise in achieving mean-field behavior. We employ finite-size scaling in order to extract critical exponents with good accuracy. We conclude that relevant structural mechanisms responsible for mean-field behavior cannot be justified in realistic models of the cortex. However, strong dynamical noise, which can have realistic justifications, always leads to mean-field behavior regardless of the underlying structure. Our work provides a different (dynamical) origin than the conventionaly accepted (structural) mechanisms for mean-field behavior in neuronal avalanches.
\end{abstract}
%%%-----------------------------------------------------------------------------

% PACS, ?the Physics and Astronomy Classification Scheme.use showpacs class
%\keywords{}%Use showkeys class option if keyword
                              %display desired

\maketitle
%--------------------------------------------------------------------------------
\section{I. Introduction}
The critical brain hypothesis is by now well supported by experimental \cite{Plenz,PT,BP1,BP2,FIBSLD,PTLNCP,TBFC,SACHHSCBP,HTBC,C} as well as theoretical studies \cite{Plenz,DanteC,BN,BR,LHG,MMKN,APH}. It is believed that the brain exists in an intricate balance between ordered and disordered states much like the standard critical point of a continuous phase transition. Such a meta-stable state is believed to underlie many novel properties of the brain including variability and/or adaptability \cite{AH}, efficient information processing \cite{Beggs}, transmission and storage of information \cite{SYYRP}, maximum sensitivity to sensory inputs \cite{SYPRP,KC}, among others. One of the most important experimental evidence in support of the critical brain hypothesis is the observation of resting state neuronal avalanches whose discovery was inspired by models of self-organized criticality (SOC) where small perturbations can lead to a wide range of events (avalanches) which exhibit scale-invariant statistics, a hallmark of a system poised at the critical point \cite{BTW1,BTW2}. Neuronal avalanches are observed in cortical slice cultures of rat cortex \cite{BP1,BP2} and also the spontaneous cortical activity of awake monkeys \cite{PTLNCP} using electrode arrays recording, as well as the resting fMRI \cite{TBFC}, and MEG \cite{SACHHSCBP} recording over the entire human cortex.

It is believed that resting state neuronal avalanches are well-modeled by threshold dynamics of SOC, where a local instability will propagate through the system via local connections to other threshold elements \cite{BTW1,BTW2,B,P}. The statistics of such events known as avalanches show scale-invariant behavior ($P(x)\sim x^{-\tau_{x}}$) both in size ($x=s$) and duration ($x=d$). This has been well-established both in experimental studies of neuronal avalanches and numerical solutions of various sandpile models. However, a certain important question remained unanswered. All experimental results show $\tau_{s}\approx 3/2$ and $\tau_{d}\approx 2.0$ consistent with a binary branching process which is the mean-field solution of sandpile models \cite{ZLS,A}, but inconsistent with actual values of all such models which exhibit exponents significantly smaller than such mean-field values when put on a two dimensional lattice with local connectivities. One might suspect that the high average connectivity in cortical neurons is the reason for such mean-field behavior. Another possible mechanism is the possibility of long-range connections which can lead to small-world effect in such networks with subsequent mean-field exponents \cite{BS}. A more recent and \textit{dynamical} (as opposed to \textit{structural}) mechanism for observation of mean-field exponents is synaptic noise in dynamics of connectivity between neurons \cite{AMIN}. In this work we propose to find which one of these mechanisms is more relevant in real cortical networks by studying properties of SOC under such circumstances.

It is worthwhile to point out that criticality of neuronal dynamics has been at times a controversial issue \cite{BT,HG}. There are authors who have questioned the authenticity of neuronal avalanches \cite{Destexhe}. Furthermore, there are certain issues associated with reliability of the reported exponents in various experimental set-ups. For example, in the original experiments of Beggs and Plenz \cite{BP1} the finite (and admittedly small) size of the multi-electrode arrays used to record the avalanches and subsequently extract the exponents, make the reported exponents somewhat unreliable. Nonetheless, the wealth of experimental as well as theoretical studies published in recent years have gone far in providing a general picture where critical dynamics is generally believed to be a fundamental property of neuronal dynamics. Our intention here is to take the mean-field like behavior of neuronal avalanches as a given, and seek to find the dominant mechanism which may lead to such a behavior. We also note that the general topic of \lq\lq structure vs. dynamics" in neuroscience is an important topic of current interest with wider outlook than the specific topic of criticality and should therefore be of interest to the general field of neuroscience.

We therefore propose to study the stochastic parallel Zhang (SPZ) model of SOC which can easily be interpreted as a simplified model of neuronal dynamics while it succumbs to simple finite-size scaling devoid of complications associated with various other sandpile models. We consider a two-dimensional (2D) regular square lattice and consider avalanche statistics of the model by varying three parameters: the average connectivity ($K$), the ratio of random long-range connections ($q$), and the strength of dynamical synaptic noise ($\sigma$). We use finite-size scaling in order to extract critical exponents of the systems under study and look for transitions to mean-field exponents as $K$, $q$, $\sigma$ are varied from their standard values of $K=4$, $q=0$ and $\sigma=0$. We find that increasing the average connectivity does not lead to mean-field behavior as long as it remains local. In fact it does not change the critical exponents at all. On the other hand increasing $q$ will clearly change the shape of avalanche distribution functions leading to mean-field behavior for large enough $q$. However, as we will argue the values of $q$ seen in simulations are not consistent with those in the cortex. Moreover, the patterns of avalanches under such conditions seem inconsistent with those seen in real neuronal avalanches. On the other hand, the inclusion of synaptic noise will always lead to mean-field behavior regardless of the structural background on which avalanches take place. Our study therefore points to a dynamical origin as opposed to a structural origin of mean-field behavior in neuronal avalanches.

The paper is structured as follows: in section II we will motivate and discuss our model. Section III is devoted to presentation of our numerical solution of the model under various conditions. Finally, we will discuss our conclusions in section IV.

\section{II. THE MODEL}
Neuronal dynamics is a threshold dynamics, i.e. the electric
potential of the membrane of a neuron must exceed a threshold
value for that neuron to fire. Neurons integrate charges that are
gained via neuronal interactions, until their membrane potential
reaches a threshold value where they fire and interact with their
neighbors through synaptic connections. This simplistic approach
to neuronal dynamics is the same as sandpile model of SOC. In order to model the threshold dynamics of neuronal avalanches, we use a sandpile model with a
continuous local variable ($E$), known as the
stochastic parallel Zhang (SPZ) sandpile model \cite{SV}.

We define the SPZ sandpile model on a  general network with $N$ nodes where every node $i$ can interact with its neighbors $j$. The number of neighbors of the node $i$ is $K_{i}$, and a binary adjacency matrix explicitly defines the neighbors of every node. Dynamics of the SPZ model on the network starts by random driving, i.e. a node is chosen randomly and its energy, i.e. its membrane potential, increases by $\delta E $, $E\rightarrow E+\delta E$, where $\delta E$ is a randomly chosen number in the range $[0,0.25]$. This emulates a random external input to a neuron in the cortex. The driving process continues until the energy of a site reaches a threshold value, $E_{th}=1$. In this state the system is unstable, and the unstable node transfers its energy to its neighbors, using the toppling rule
\begin{equation}\label{Eq1}
E_{j}\rightarrow E_{j}+\epsilon_{j}E_{i} \hskip5pt, \hskip5pt   E_{i}\rightarrow 0
\end{equation}
 where $\epsilon_{j}$ are $K_{i}$ annealed random numbers in the range $[0,1]$, with the constraint $\sum_{j=1}^{K_{i}} \epsilon_{j}=1$, which guarantees local conservation of energy. Toppling of a site increases the energy of its neighbors and can therefore make them unstable. The new unstable sites must topple by the same toppling rule and this process continues until the time that no unstable sites remain. The totality of this relaxational process which starts with a single site instability is called an avalanche. When the system comes into a stable state, random driving starts in order to perturb the system and start a new avalanche. The balance between this slow driving (e.g. resting state) and dissipation at the open boundaries help keep the system near the critical point despite the fact that the system is continuously driven. The separation of time-scales along with the above-mentioned \textit{local} conservation are the essential ingredients which self-organize the system to a critical point where avalanche statistics show scale invariant behavior. The number of topplings (firings) in an avalanche is defined as the size ($s$), and the number of time steps as the duration ($d$) of an avalanche.

In the case of the noisy local dynamics \cite{AMIN} we change the
toppling rule to
\begin{equation}\label{Eq2}
E_{j}\rightarrow E_{j}+\epsilon_{j}E_{i}+\eta_{j} \hskip5pt , \hskip5pt E_{i} \rightarrow 0
\end{equation}
where $\eta_{j}$ is a randomly chosen annealed flat noise in the range $[-\sigma , \sigma]$ with zero mean value $\langle \eta \rangle=0$. When the noise mean value is equal to zero, the energy of the system is conserved on the average, but the strict local
conservation can be broken due to the noisy dynamics. The addition of this simple annealed noise is meant to emulate the noisy random synaptic interaction between neurons. In a previous work \cite{AMIN}, we have shown that introducing noise into the toppling rule does not destroy the
criticality of the system, but it changes the critical properties
of the system. We have shown that, with increasing $\sigma$ the critical exponents of the system will gradually increase until they reach and saturate at their mean-field values at $\sigma\approx 0.25$ \cite{AMIN}. Our previous work showed such results on regular rectangular two and three dimensional lattices of $K_{i}=2D , \forall i$. It must be noted that we use a flat noise in order to streamline our computer simulations.

In order to study the impact of the small-world effect on the
mean-field behavior of the model we use the Newman-Watts method
\cite{WN} to build a network with the small-world property. We
start with a two dimensional regular network with $N$ nodes, every
node interacts with $K$ neighbors. There exists $NK/2$ links in
the system at this step, then we start to add new random links to the system, i.e. we add a link between two
randomly chosen nodes that are not already connected. The adding of
new links continues until the ratio of the random links equals $q$.

We must note that, in the case of regular lattice if an avalanche
reaches the boundaries of the system the energy dissipates through
the boundary sites and addition of new links in the small-world
networks does not change the dissipative sites in our simulations.

Before closing this section we emphasize that the dynamics of SPZ model is a simplistic but perhaps adequate model --within the usual minimalist approach to critical systems-- of neuronal dynamics as the external inputs lead to increase of membrane potential ($E$) and thus possible firing of a random neuron which subsequently resets itself and leads to synaptic interaction with its neighboring neurons with random weights $\epsilon_{j}$ (see Eq.\ref{Eq1}). Addition of annealed noise (Eq.\ref{Eq2}) is meant to mimic the random effect of neurotransmitters available at the time of synaptic interaction. Note that the separation of time scales implicit in the SPZ model is also relevant in neuronal avalanches \cite{Beggs}. Therefore our proposed model is a self-organizing model with threshold dynamics of a continuous dynamical variable easily associated with membrane potential of a neuron. It exhibits critical behavior which can reliably be studied by finite-size scaling analysis unlike many other SOC models which exhibit multi-scaling and unusual finite-size scaling behavior \cite{P}. It also easily allows us to add synaptic noise. Another key point about the present model is that it does \textit{not} show mean-field critical exponents in its standard form ($K=4, q=0, \sigma=0$), see Table \ref{table1}. It therefore allows us to investigate in a physically meaningful way how the increase of such parameters (in a neurologically motivated manner) will lead to the mean-field behavior seen in the experiments. Clearly, a model which generically produces mean-field exponents cannot afford such an analysis.

\section{III. RESULTS}

In SOC systems like the sandpile models where the system is
critical, and consequently scale invariant, probability
distribution functions of the size and duration of avalanches,
usually obey the simple scaling ansatz $P(x)\sim x^{-\tau_{x}}
f(x/L^{\beta_{x}})$ \cite{P}, in which $x$ can be either $s$ or
$d$, and $f(x/L^{\beta_{x}}) $ is a universal cutoff function
that arises from finite-size effects, where $L$ is the linear size
of the system and $\beta_{x}$ is the finite-size exponent. In
order to confirm scale invariance and also to find the critical
exponents $\tau_{x}$ and the finite-size exponents $\beta_{x}$, we
use a simple finite-size scaling method. In this method, if $P(x)$
is rescaled as $P(x)\rightarrow x^{\tau_{x}}P(x)$, and
$x\rightarrow x/L^{\beta_{x}}$, then plots of the rescaled
variables, i.e. $x^{\tau_{x}}P(x)$ versus $x/L^{\beta_{x}}$, must
collapse into a single universal curve for different values of $L$
\cite{VP}. In some cases, like the upper critical
dimension of SPZ sandpile model \cite{AMIN}, noisy local dynamics
in SPZ model \cite{AMIN}, stochastic Manna model in two
dimensions \cite{DC}, BTW model on scale free networks
\cite{GLKK} and its upper critical dimension (D=4) \cite{LU},
there exists a logarithmic correction to the scaling ansatz, and
it is written in the form $P(x)\sim x^{-\tau_{x}}
[\ln(x)]^{\gamma_{x}} f(x/L^{\beta_{x}})$, where we need an
additional exponent, $\gamma_{x}$. In the presence of the
logarithmic correction, $P(x)$ must be rescaled as
$P(x)\rightarrow x^{\tau_{x}}[\ln(x)]^{-\gamma_{x}} P(x)$ to get
good scaling collapses.

\begin{figure}[!htbp]
	\begin{center}
		\includegraphics[width=0.4\textwidth]{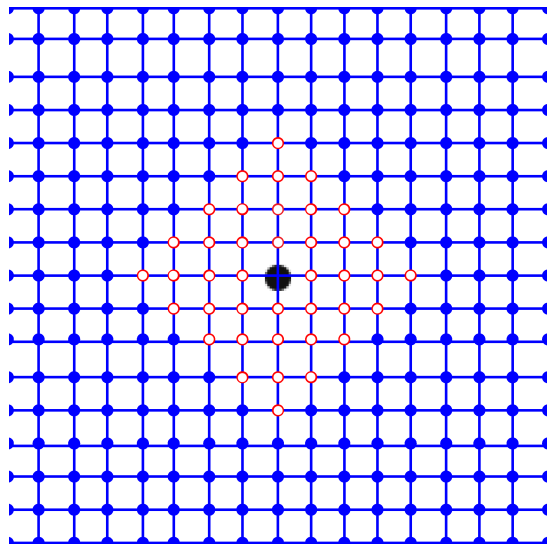}
	\end{center}
	\caption{(Color online) The sites shown in red open circles are the $40$ neighbors of the central site colored in black on a regular two dimensional lattice.}
	\label{fig1}
\end{figure}

Theoretically, mean-field solutions are exact for a system with
all to all connectivity, or dimensionality greater than the upper
critical dimension where effectively fluctuations become irrelevant \cite{Krdarf}. In SOC systems it has been shown that mean-field behavior is caused by topological as well as dynamical
properties of the model. For example, the BTW sandpile model on scale-free
networks exhibits mean-field behavior for the exponent of the
degree distribution function that are greater or equal to three, i.e. $P(K)\sim K^{-\alpha}$, $\alpha\geq 3$ \cite{GLKK}. This model also shows mean-field behavior on random graphs \cite{BON}, and small-world networks with the probability
of long-range connections greater or equal to $q=0.1$ \cite{BS}.
Also as mentioned above, it has been shown that the noisy local dynamics in the SPZ
sandpile model leads to mean-field results for strong enough noise
level \cite{AMIN}. Topological properties of neural networks, like
high connectivity and small-world effects, are generally
considered as the reasons for observing the mean-field behavior in
neuronal avalanches, where the dynamics of the system can be
mapped to a random-neighbor theory or a branching process, that
are known to be mean-field solutions for the SOC sandpile models \cite{Plenz,BP2}. By analyzing the SOC models with high local connectivity, small-world effects and noisy
local dynamics, we can find out which property may play the
main role in observing mean-field behavior in actual neuronal
avalanches.

\begin{figure}[!htbp]
	\begin{center}
		\subfigure[]{\includegraphics[width=0.48\textwidth,height=0.45\textwidth]{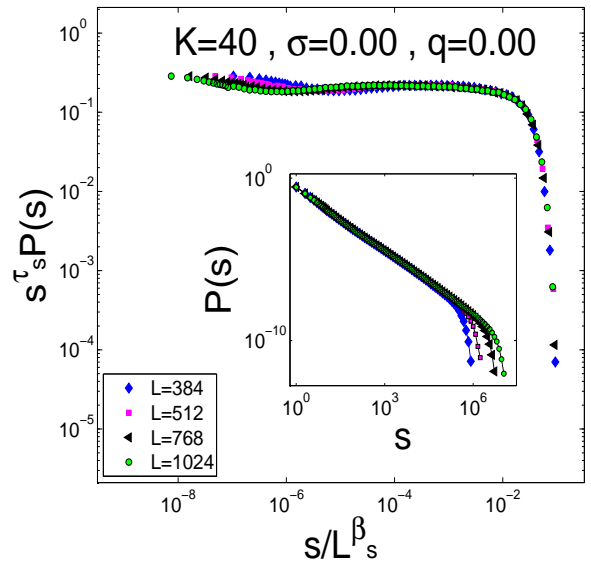}\label{fig2a}}
		\subfigure[]{\includegraphics[width=0.48\textwidth,height=0.45\textwidth]{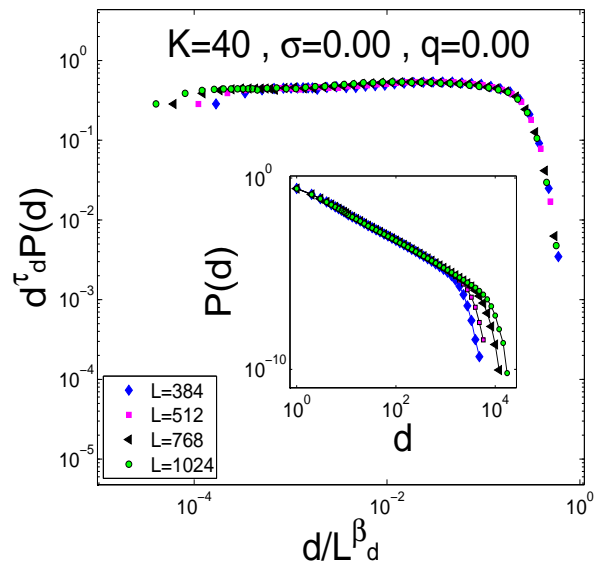}\label{fig2b}}
	\end{center}
	\caption{Finite-size-scaling collapse for (a) size, and (b)
		duration of avalanches for two dimensional SPZ model with $K=40$.
		Linear system sizes are $L=384, 512, 768, 1024$. The
		exponents obtained from the collapses are reported in Table \ref{table1}. Insets show the uncollapsed data.}
	\label{fig2}
\end{figure}

\begin{table}[!htbp]
	\begin{tabular}{l|c|c|cc|ccr}
		$K$ & $q$ & $\sigma$ & $\tau_{s}$ & $\beta_{s}$ & $\tau_{d}$ & $\gamma_{d}$ & $\beta_{d}$ \\
		\hline
		4  & 0.00 & 0.00 & 1.28(1) & 2.76(2) & 1.50(1) & 0.00 & 1.53(1)\\
		40 & 0.00 & 0.00 & 1.27(1) & 2.70(2) & 1.49(1) & 0.00 & 1.46(2)\\
		\hline
		4  & 0.10 & 0.00 & 1.50(1) & 2.00(2) & 2.00(1) & 0.50(5) & 1.00(2)\\
		40 & 0.10 & 0.00 & 1.50(1) & 2.00(2) & 2.00(1) & 0.50(5) & 1.00(2)\\
		\hline
		40 & 0.00  & 0.25 & 1.49(2) & 4.0(1) & 2.00(2) & 0.70(8) & 2.00(5)\\
		40 & 0.005 & 0.25 & 1.51(2) & 2.20(5) & 2.01(2) & 0.60(8) & 1.00(5)\\
	\end{tabular}
	\caption{Exponents obtained from various finite-size scaling collapses in this work. For details see the ensuing text and figures.}
	\label{table1}
\end{table}

\begin{figure}[!htbp]
	\begin{center}
		\subfigure[]{\includegraphics[width=0.48\textwidth,height=0.45\textwidth]{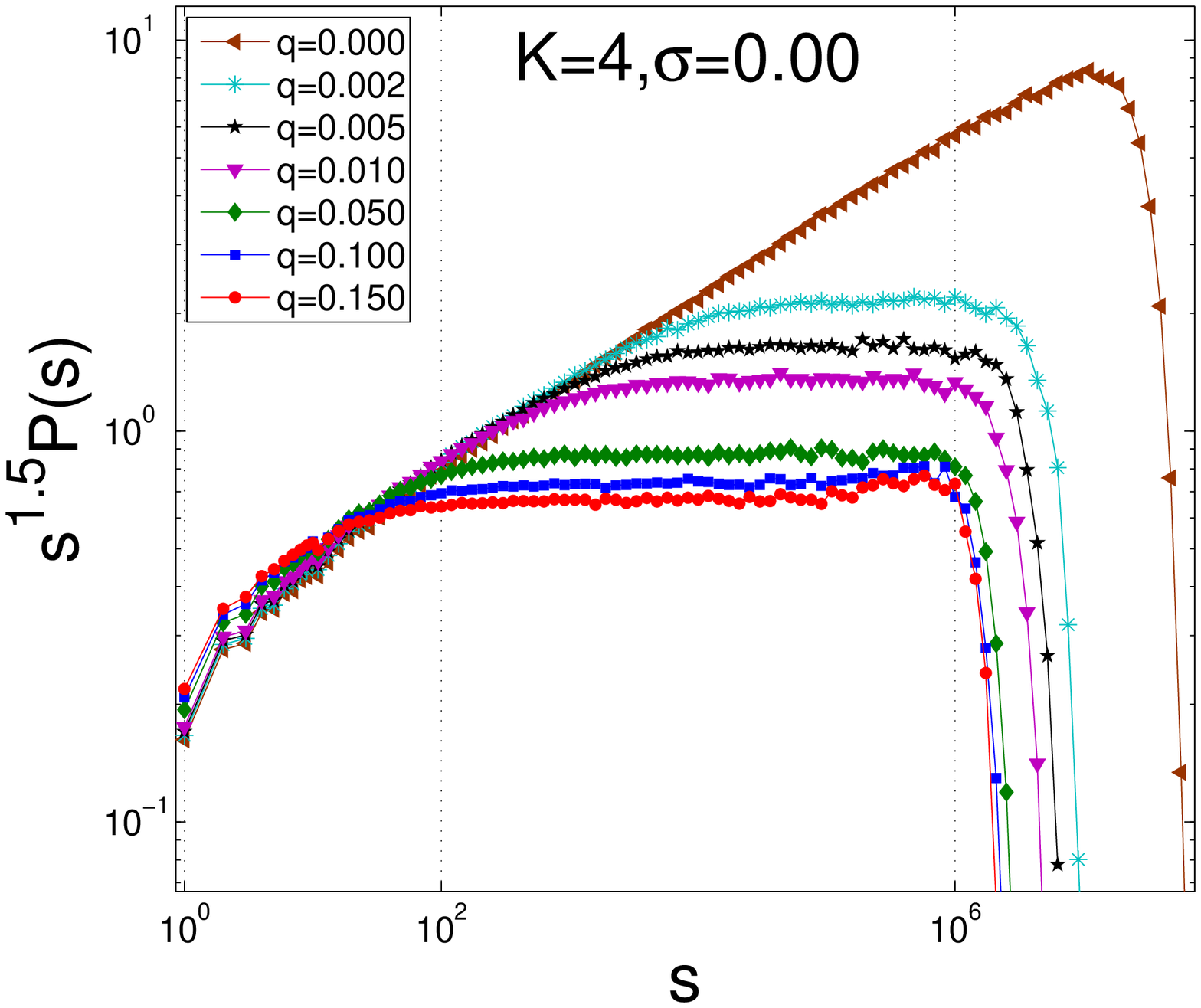}\label{fig3a}}
		\subfigure[]{\includegraphics[width=0.48\textwidth,height=0.45\textwidth]{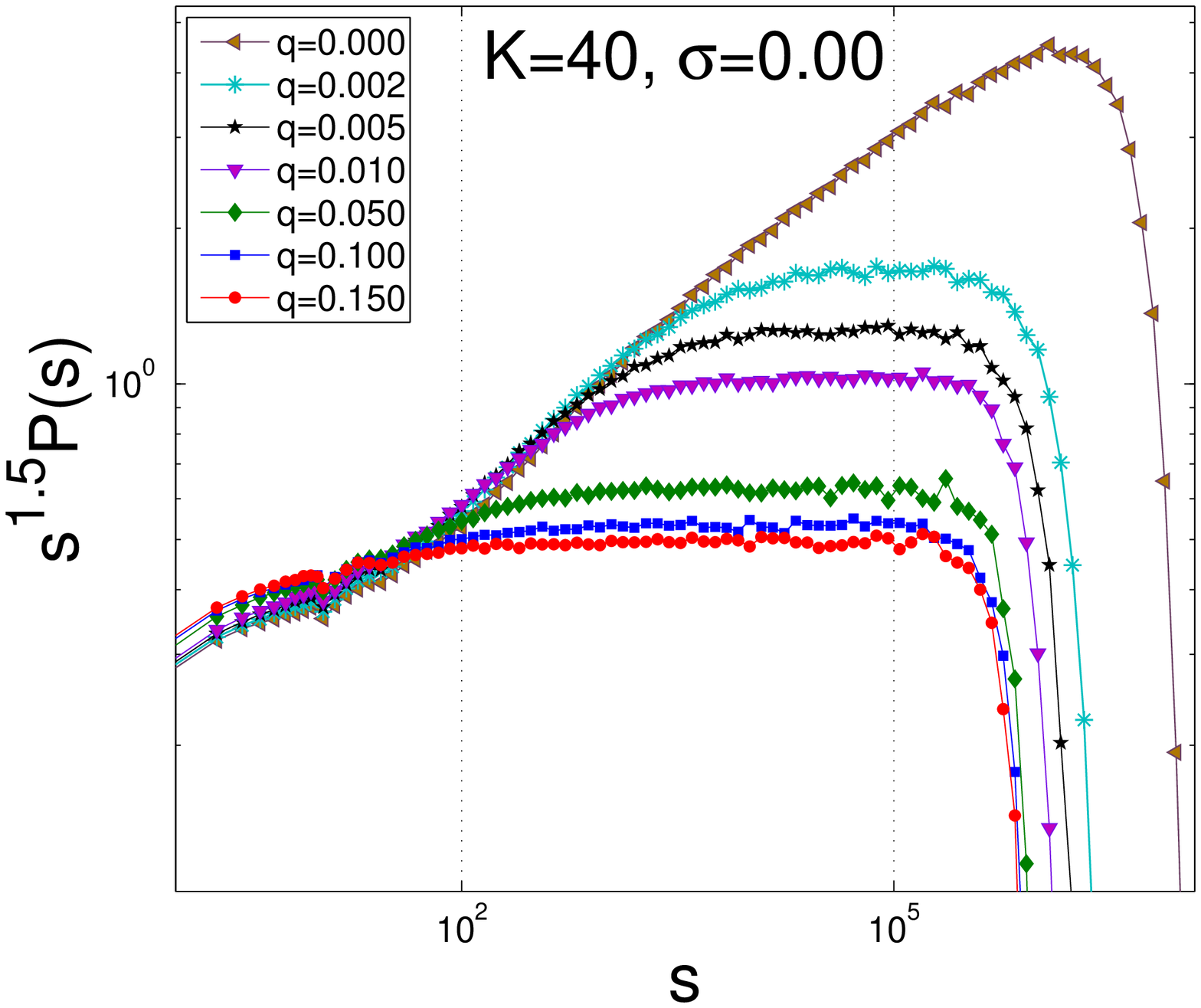}\label{fig3b}}
	\end{center}
	\caption{The $y$ axis of the probability
		distribution functions of the size of avalanches is rescaled by $P(s)\rightarrow s^{1.5}P(s)$ where a horizontal flat part of the plots is an indication of the mean field exponent ($\tau_{s}=1.5$). The systems are SPZ model with (a) $K=4$, and (b) $K=40$, on Newman-Watts networks with different values of $q$. Linear system size is $L=1024$. Comparing panels (a) and (b) we can conclude that mean-field behavior of the system depends crucially on the value of ($q$) and is not significantly effected by the value of $K$.}
	\label{fig3}
\end{figure}

\begin{figure*}[!htbp]
	
	\includegraphics[width=0.90\textwidth,height=0.8\textwidth]{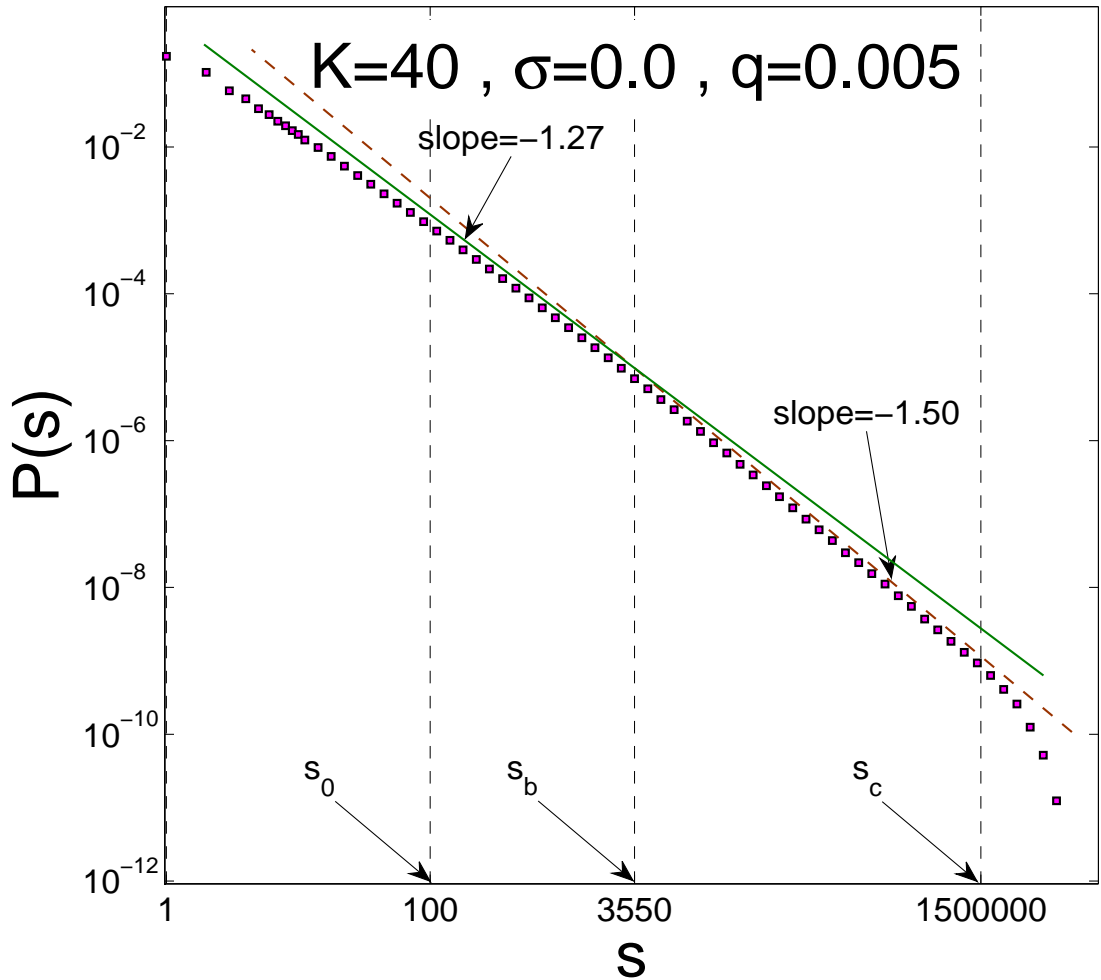}
	\caption{Behavior of the SPZ model on a Newman-Watts network with $L=1024$, $q=0.005$, $\sigma=0.00$ and $K=40$ is analyzed. The probability distribution function of avalanche sizes $P(s)$ is plotted versus $s$ (square symbols). We use a simple regression analysis in which a line with slope $-1.27$ (solid line), corresponding to simple $2D$ behavior, and a line with slope $-1.5$ (dashed line), corresponding to mean-field behavior, are fitted to the probability distribution function over the possible ranges of data. It is clear that in the range $s<s_{0}\approx 100$ the curve exhibits a different behavior from the two above-mentioned cases \cite{foot1}. The slope of the curve in the range $s_{0}<s<s_{b}\approx 3550$  is equal to $-1.27$, and in the range $s_{b}<s<s_{c}\approx 1.5\times 10^{6}$ the mean-field behavior is observed. A cutoff in the curve which is due to finite-size effects is seen for $s>s_{c}$. The value of $s_{b}$ is obtained by finding the crossing point of the dashed and the solid line. }
	\label{fig4}
\end{figure*}

\begin{figure}[!htbp]
	\subfigure[]{\includegraphics[width=0.48\textwidth,height=0.45\textwidth]{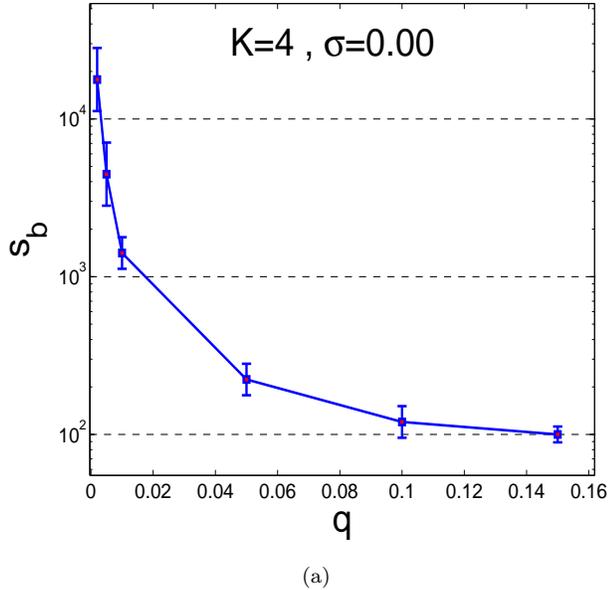}\label{fig5a}}
	\subfigure[]{\includegraphics[width=0.48\textwidth,height=0.45\textwidth]{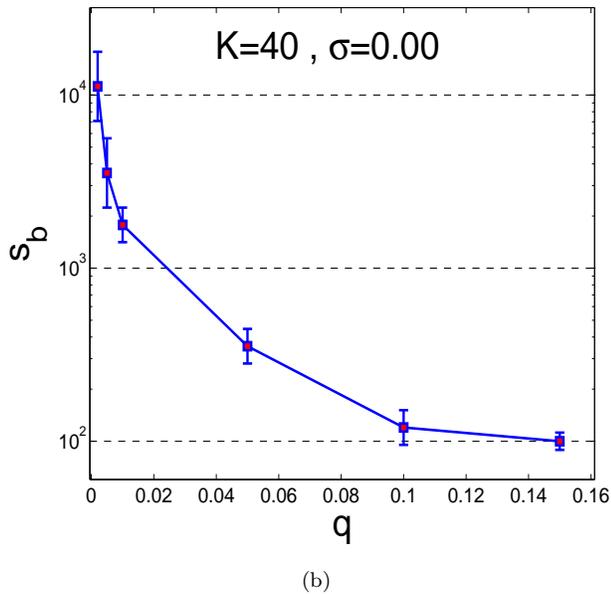}\label{fig5b}}
	\caption{The systems are SPZ model with (a) $K=4$, and (b) $K=40$, on Newman-Watts networks. The linear system size is $L=1024$. Both plots show $s_{b}$ as a function of $q$ that saturates at the value of $s_{0}\approx 10^2$ for $q\gtrsim 0.1$. The methods of obtaining $s_{0}$ and $s_{b}$ is explained in Fig.\ref{fig4}}
	\label{fig5}
\end{figure}

\begin{figure}[!htbp]
	\subfigure[]{\includegraphics[width=0.48\textwidth,height=0.45\textwidth]{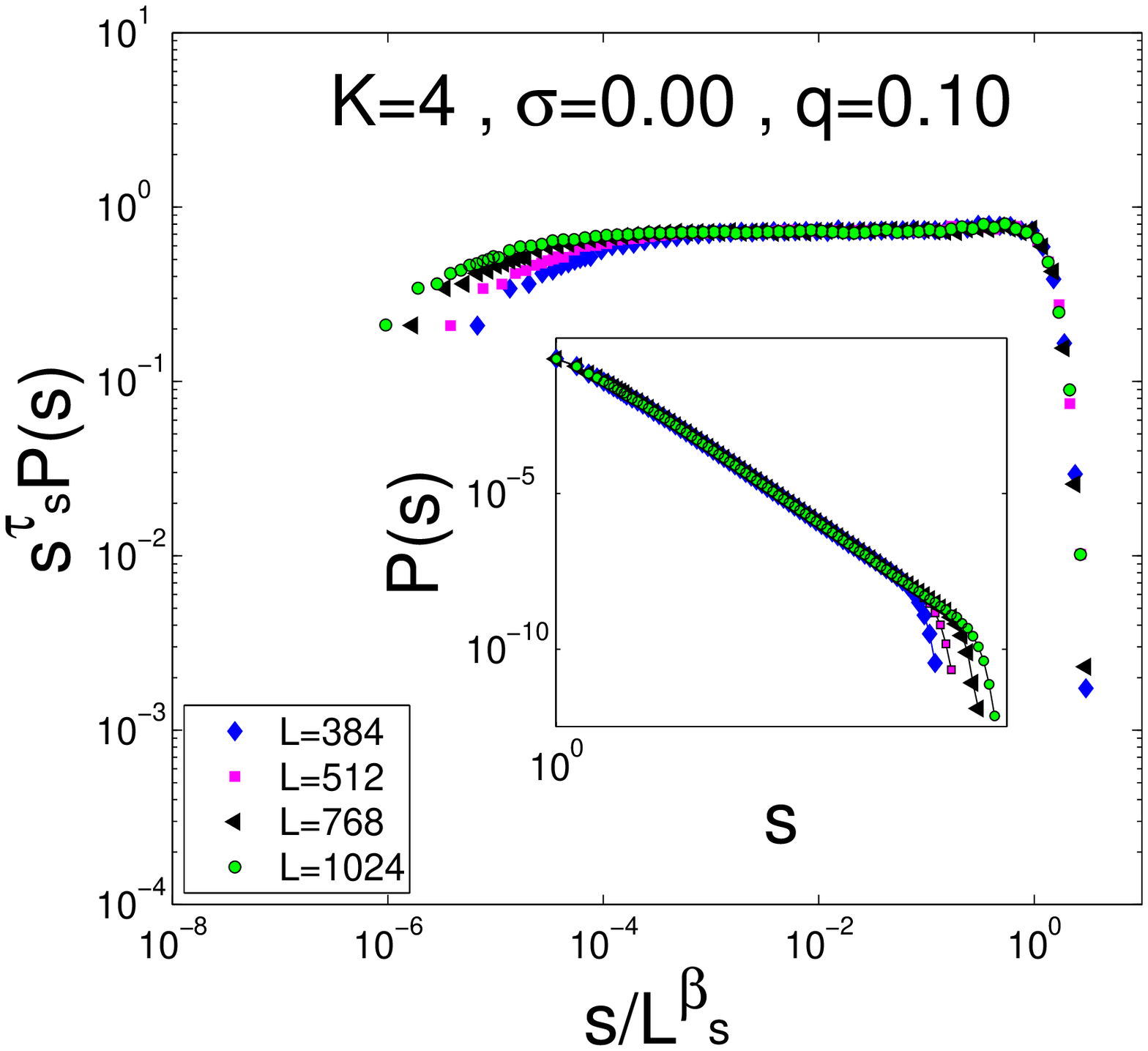}\label{fig6a}}
	\subfigure[]{\includegraphics[width=0.48\textwidth,height=0.45\textwidth]{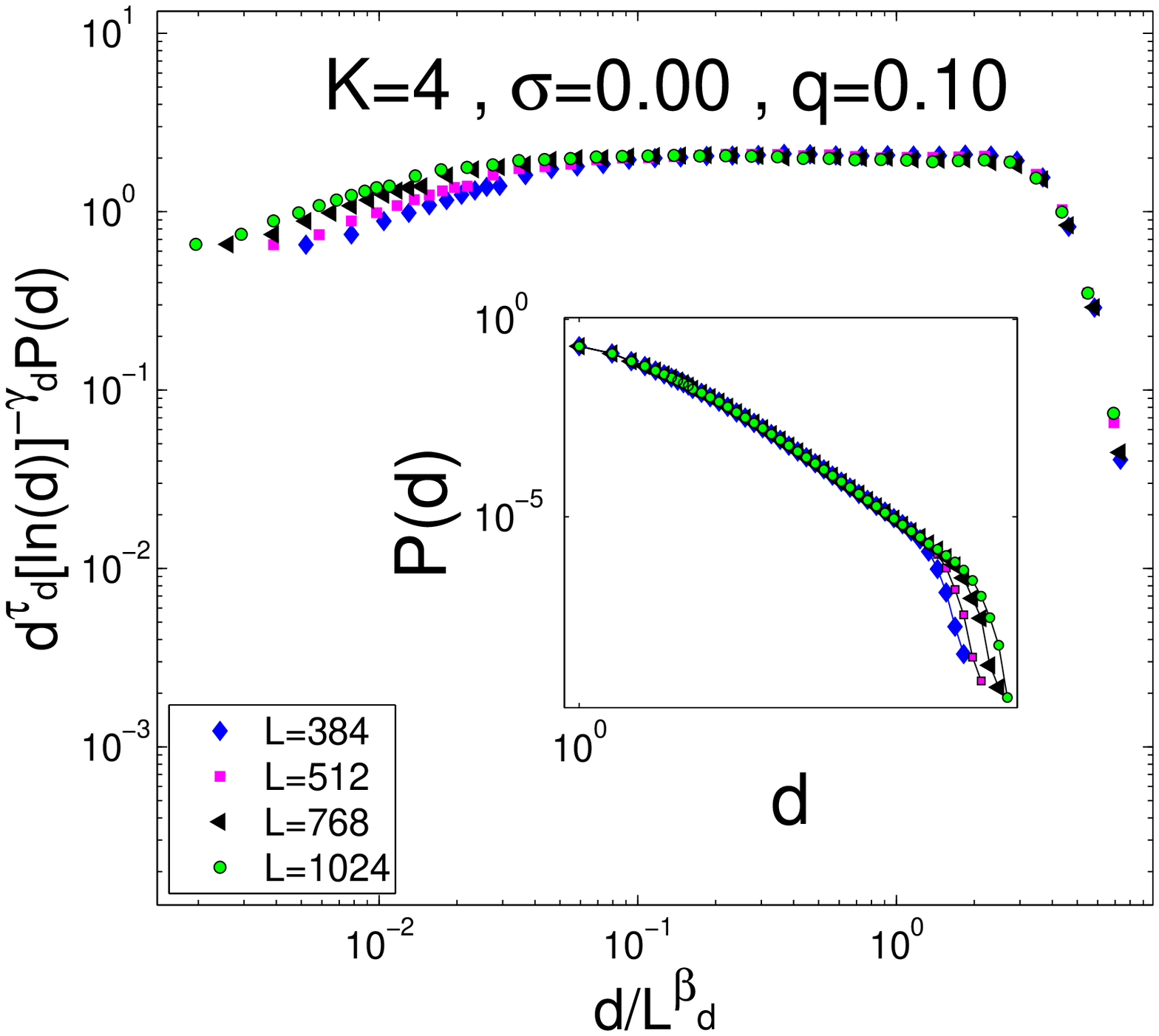}\label{fig6b}}
	\caption{Finite-size-scaling collapse for (a) size, and (b)
		duration of avalanches for the SPZ model on the Newman-Watts
		network with $K=4$ and $q=0.1$. Linear system sizes are $L=384,
		512, 768, 1024$. The exponents obtained from the
		collapses are reported in Table \ref{table1}. Note that for duration logarithmic correction results in a better collapse. Insets show the uncollapsed data.} \label{fig6}
\end{figure}

\begin{figure}[!htbp]
	\begin{center}
		\subfigure[]{\includegraphics[width=0.48\textwidth,height=0.45\textwidth]{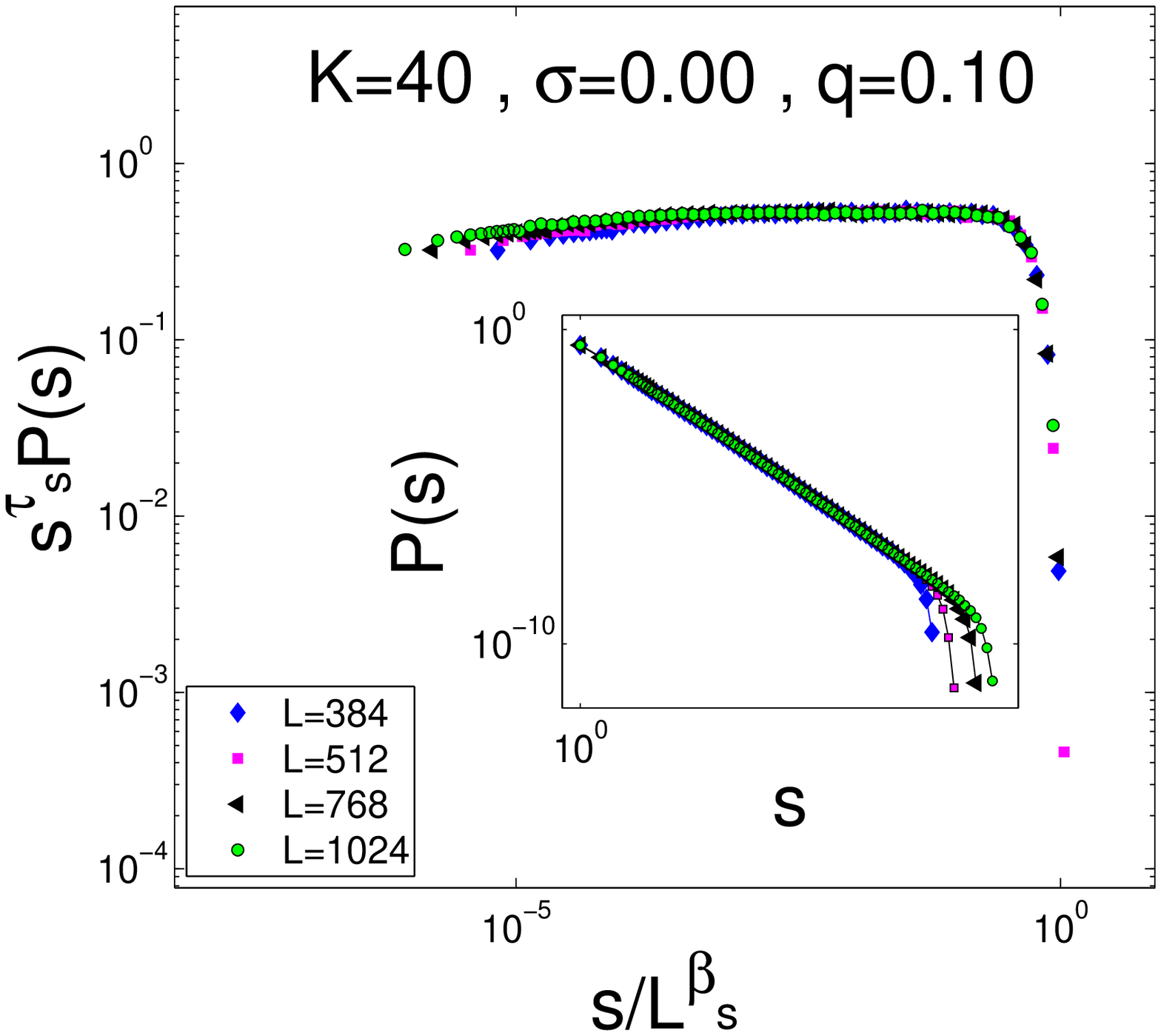}\label{fig7a}}
		\subfigure[]{\includegraphics[width=0.48\textwidth,height=0.45\textwidth]{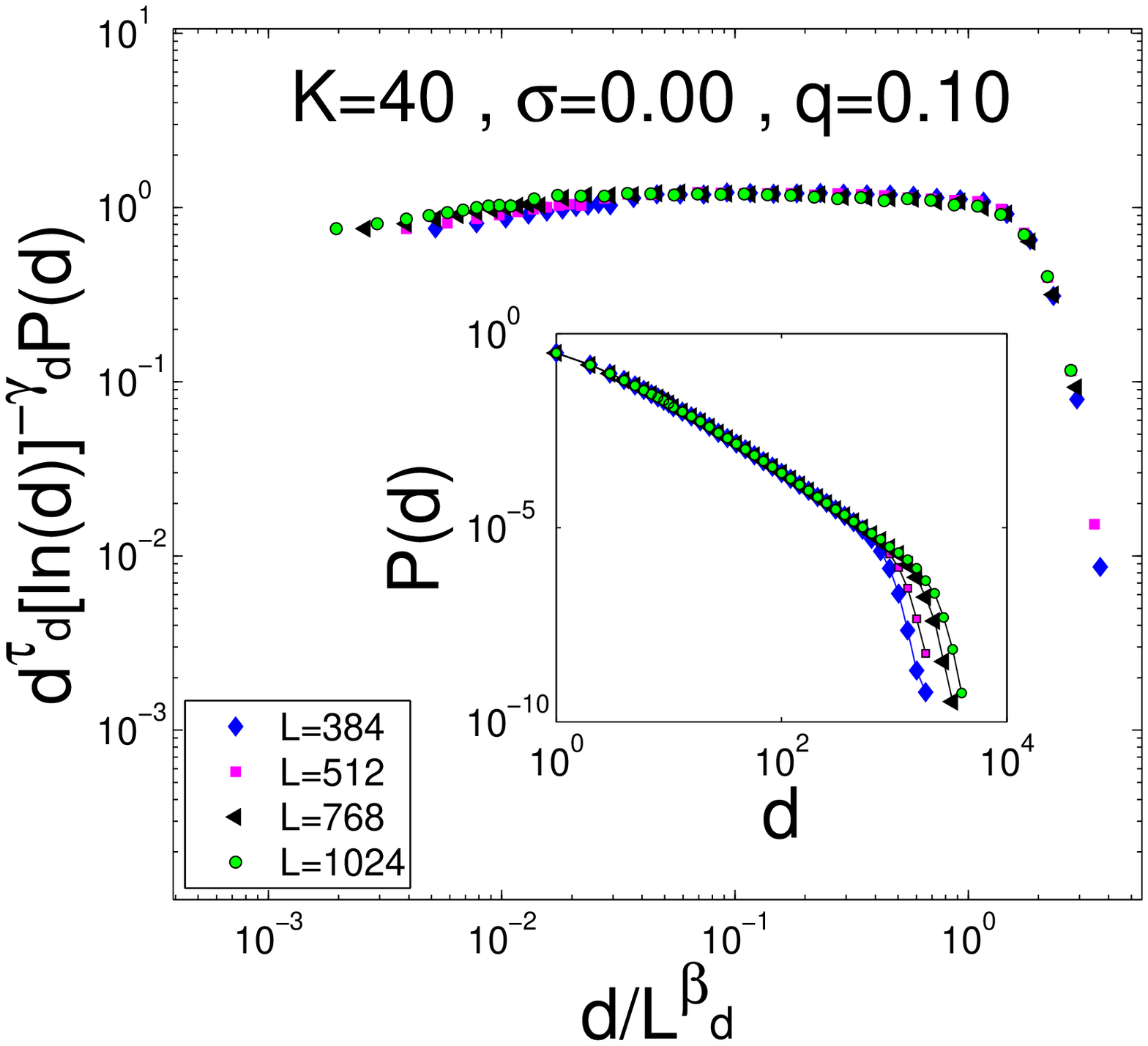}\label{fig7b}}
	\end{center}
	\caption{Finite-size-scaling collapse for (a) size, and (b)
		duration of avalanches for the SPZ model on the Newman-Watts
		network with $K=40$ and $q=0.1$. Linear system sizes are $L=384,
		512, 768, 1024$. The exponents obtained from the
		collapses are reported in Table \ref{table1}. Note that for duration logarithmic correction results in a better collapse. Insets show the uncollapsed data.} \label{fig7}
\end{figure}

\begin{figure}[!htbp]
	\begin{center}
		\subfigure[]{\includegraphics[width=0.48\textwidth,height=0.45\textwidth]{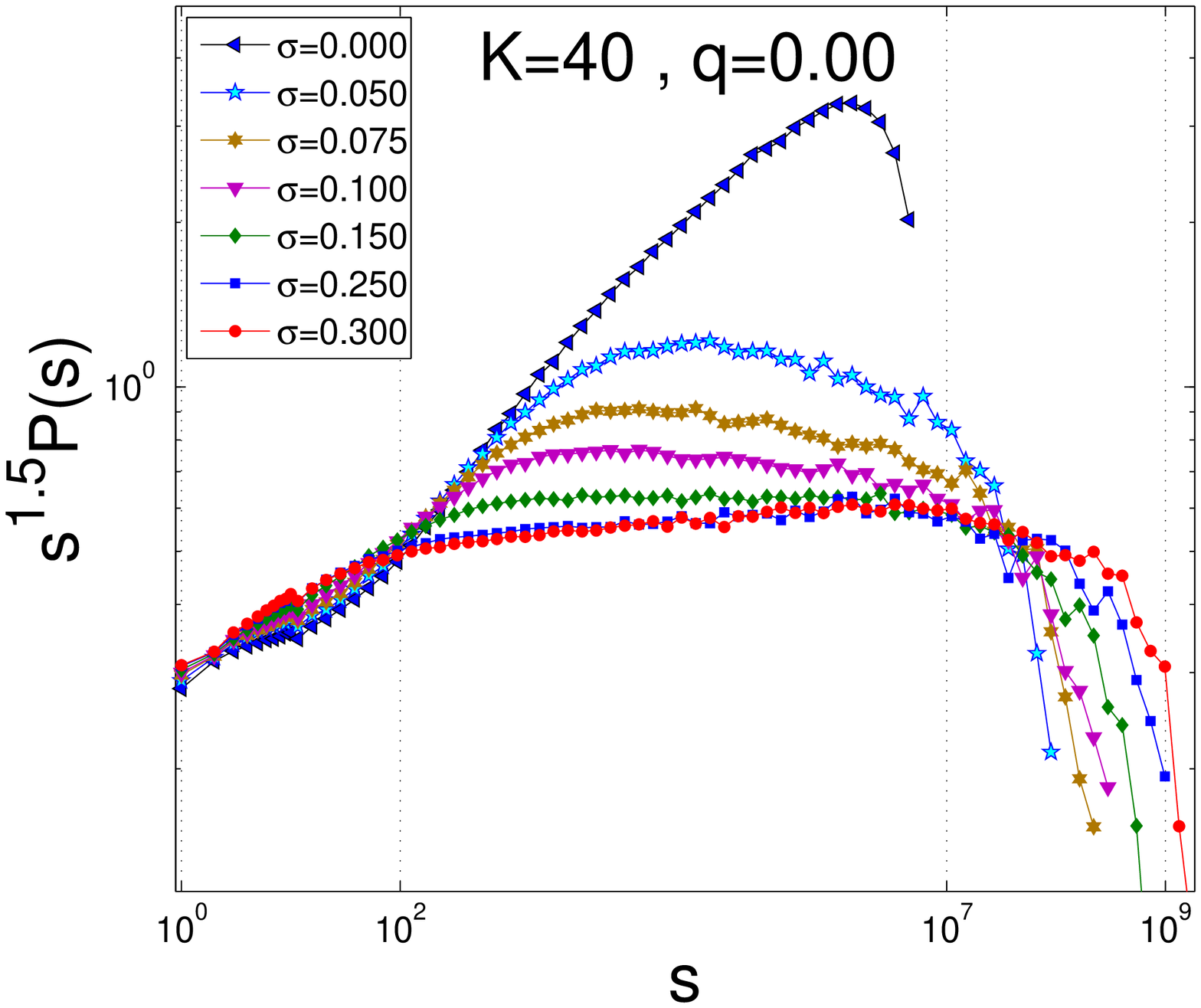}\label{fig8a}}
		\subfigure[]{\includegraphics[width=0.48\textwidth,height=0.45\textwidth]{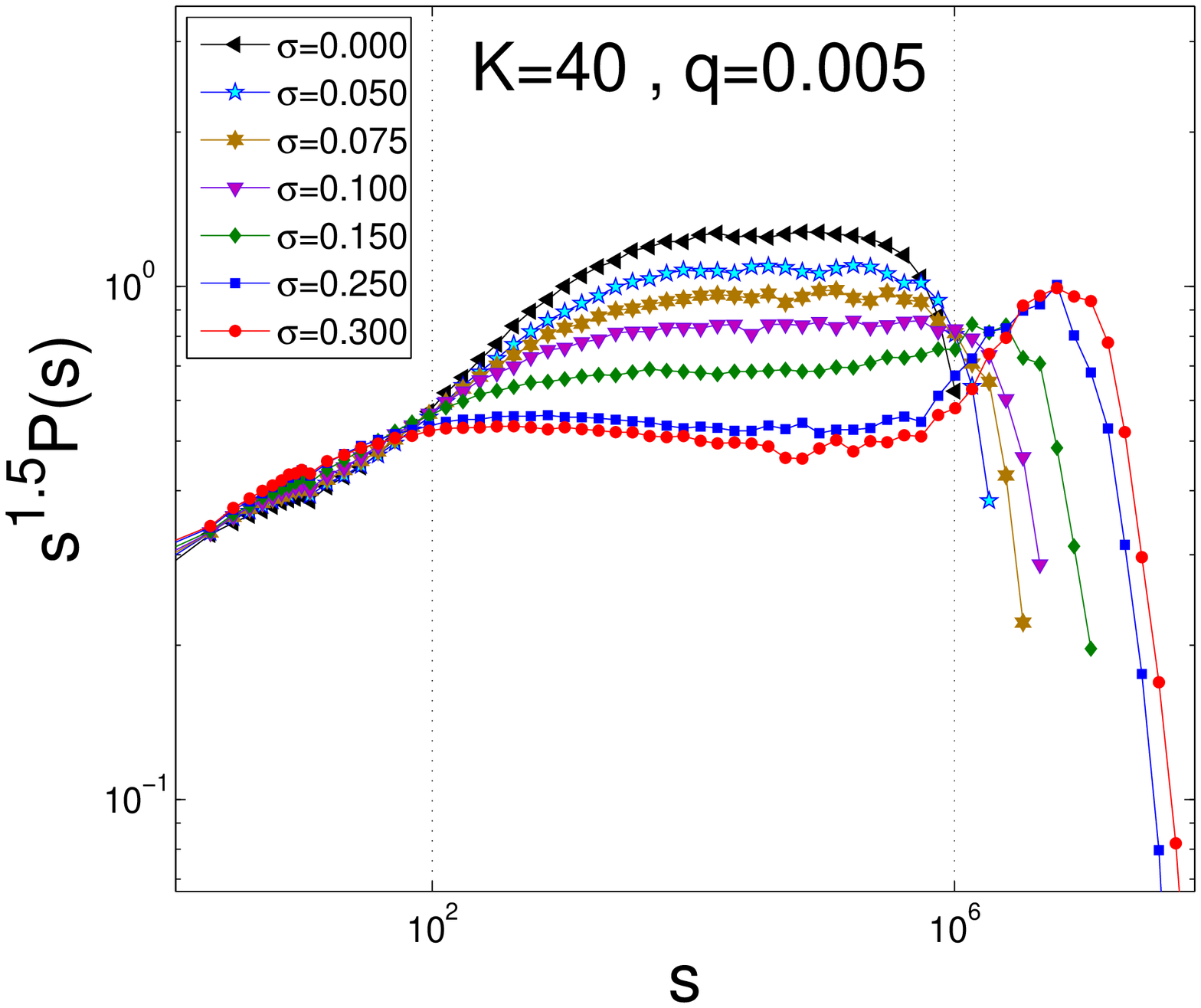}\label{fig8b}}
	\end{center}
	\caption{The $y$ axis of the probability
		distribution functions of the avalanche sizes are rescaled by $P(s)\rightarrow s^{1.5}P(s)$, where a horizontal flat part of the plots is an indication of the mean field exponent $\tau_{s}=1.5$ (slight deviation from mean-field exponent is due to finite-size effects). The systems are SPZ models with $K=40$ (a) on a two dimensional regular network, and (b) Newman-Watts network with $q=0.005$. Linear system size corresponding to both plots is $L=1024$. Comparing panels (a) and (b) we can conclude that, in both cases of small-world and regular networks, mean-field behavior over the entire range of data emerges at the same value of the noise level ($\sigma\approx 0.25$).} \label{fig8}
\end{figure}

\begin{figure}[!htbp]
	\begin{center}
		\subfigure[]{\includegraphics[width=0.48\textwidth,height=0.45\textwidth]{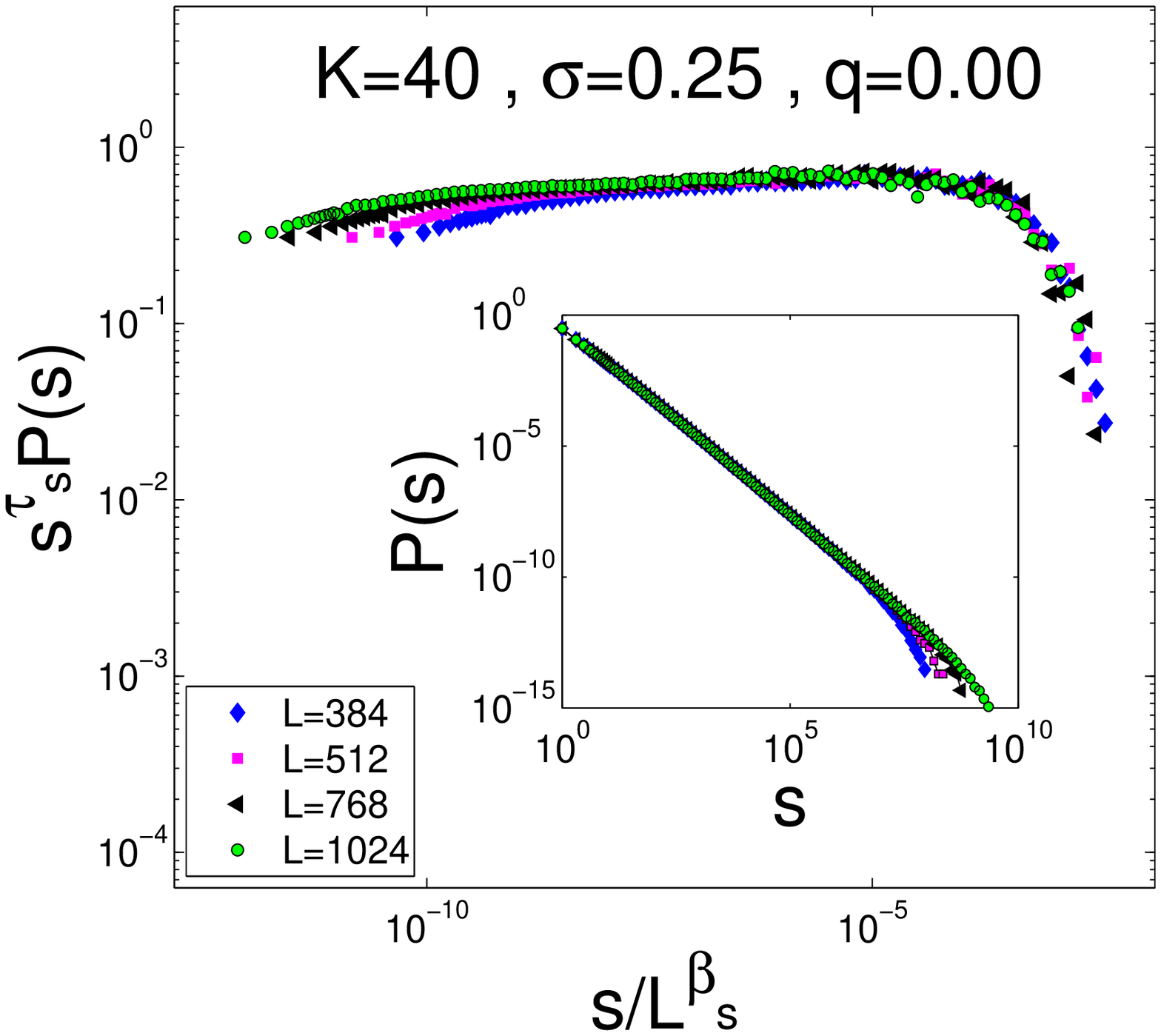}\label{fig9a}}
		\subfigure[]{\includegraphics[width=0.48\textwidth,height=0.45\textwidth]{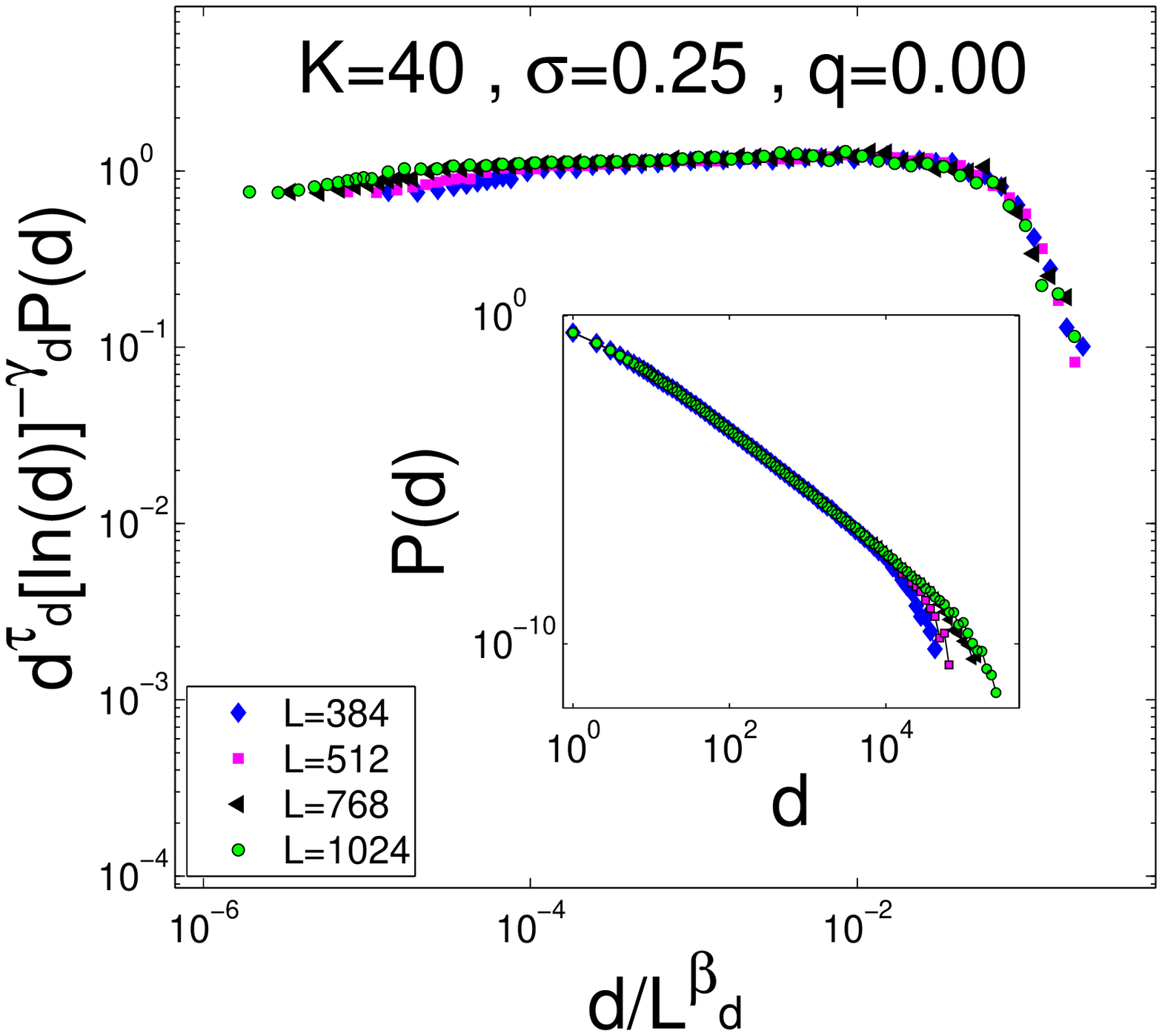}\label{fig9b}}
	\end{center}
	\caption{Finite-size-scaling collapse for (a) size, and (b)
		duration of avalanches for two dimensional Noisy SPZ model with
		$K=40$, and $\sigma=0.25$. Linear system sizes are $L=384, 512,
		768, 1024$. The exponents obtained from the
		collapses are reported in Table \ref{table1}. Note that for duration logarithmic correction results in a better collapse. Insets show the uncollapsed data.} \label{fig9}
\end{figure}

\begin{figure}[!htbp]
	\begin{center}
		\subfigure[]{\includegraphics[width=0.48\textwidth,height=0.45\textwidth]{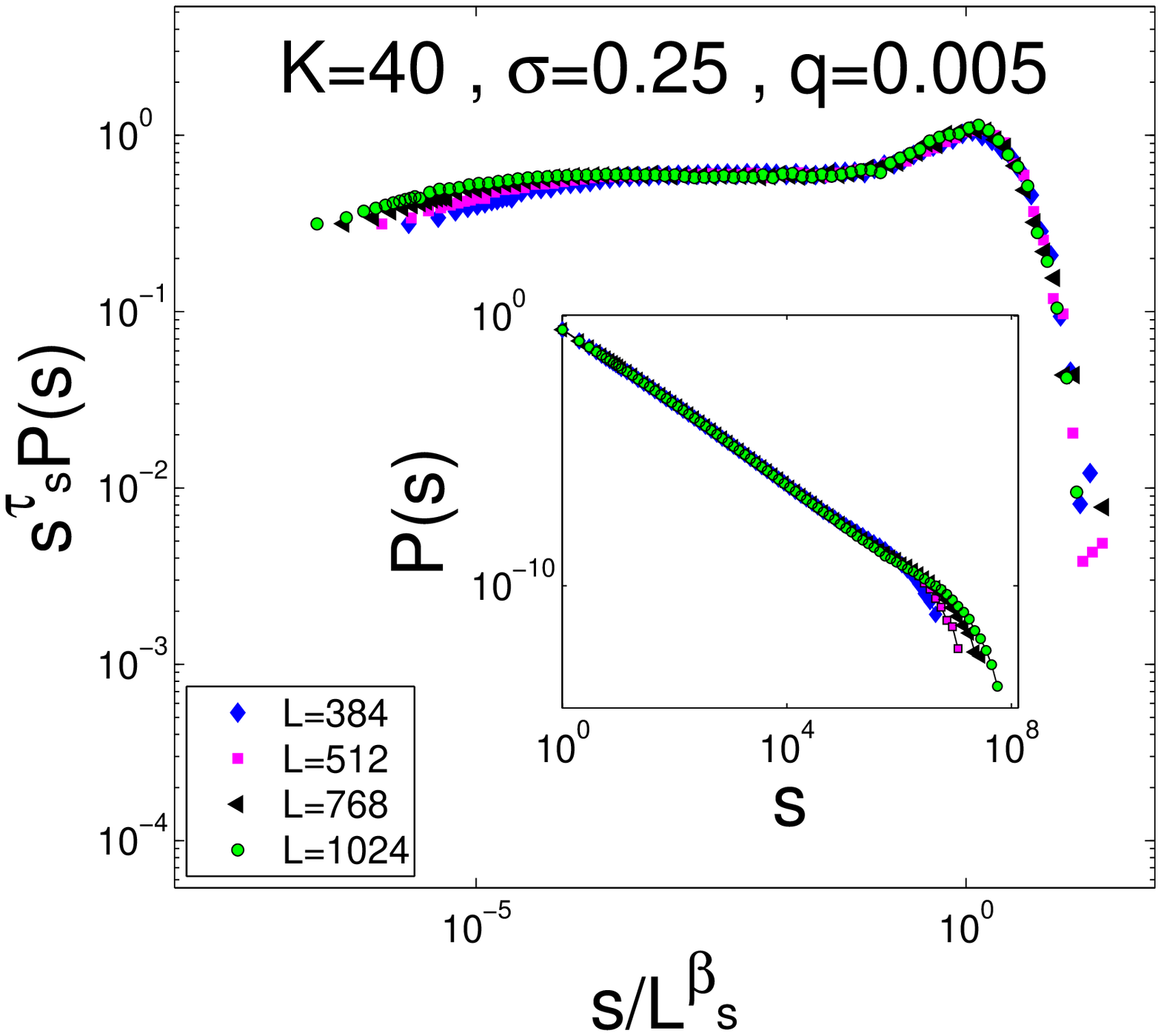}\label{fig10a}}
		\subfigure[]{\includegraphics[width=0.48\textwidth,height=0.45\textwidth]{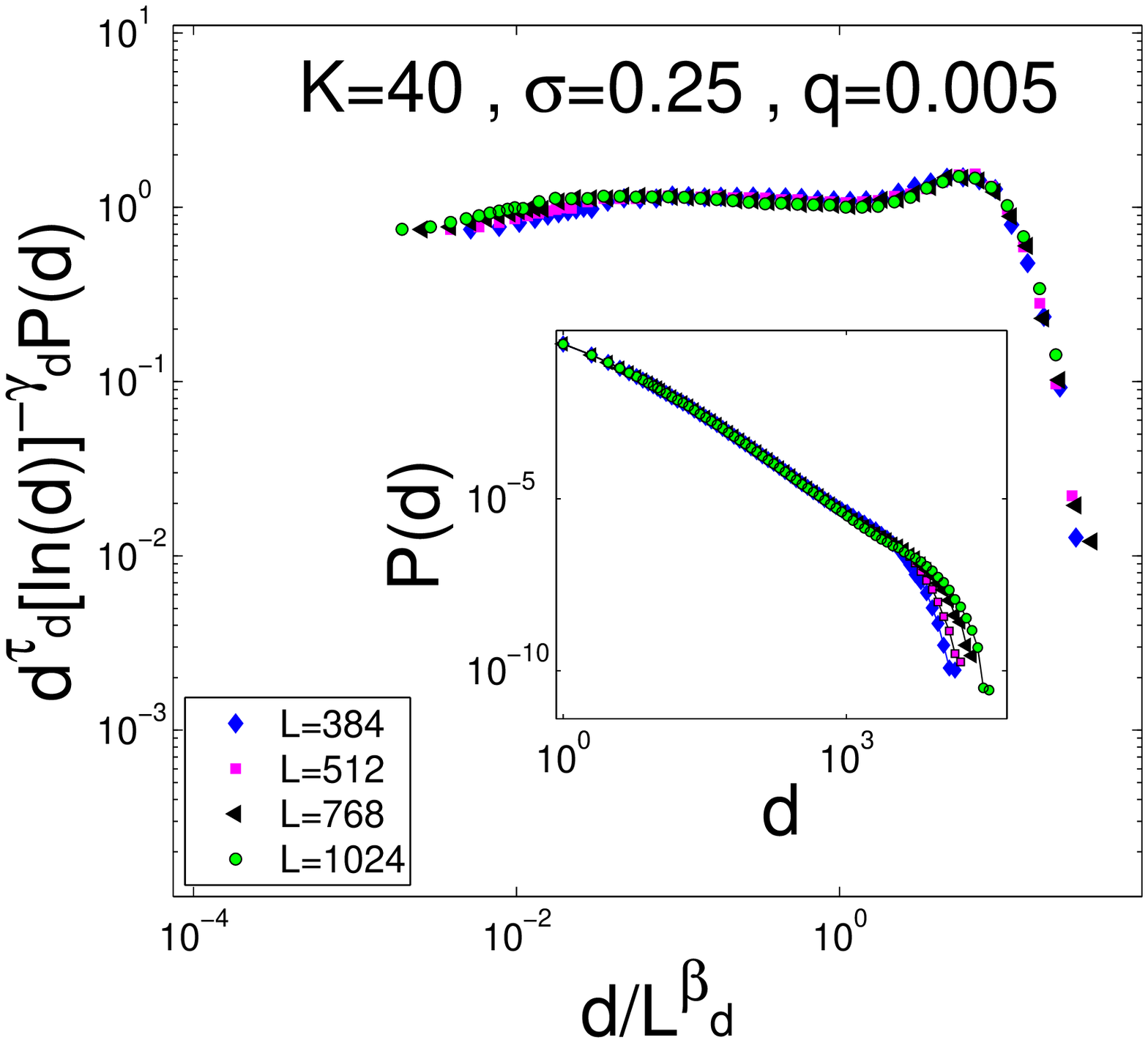}\label{fig10b}}
	\end{center}
	\caption{Finite-size-scaling collapse for (a) size, and (b) duration of avalanches for the noisy SPZ model on the Newman-Watts network with $K=40$, $q=0.005$, and $\sigma=0.25$. Linear system sizes are $L=384, 512, 768, 1024$. The exponents obtained from the collapses are reported in Table \ref{table1}. Note that for duration logarithmic correction results in a better collapse. Insets show the uncollapsed data.}
	\label{fig10}
\end{figure}

The SPZ model in two dimensions, with nearest neighbor interactions ($K=4$), exhibits good finite-size scaling collapses for size and duration of avalanches with the critical exponents of $\tau_{s}=1.28$ and $\tau_{d}=1.50$ \cite{SV}. We therefore propose to use finite-size scaling methods in order to monitor the change of these critical exponents as various parameters of our model are increased from their standard value of $K=4$, $q=0$, $\sigma=0$. We are particularly interested in the conditions under which such exponents reach their mean-field values of $\tau_{s}=3/2$ and $\tau_{d}=2.0$

To investigate the effects of increasing local connectivity, we have simulated the SPZ model for a system with high local connectivity ($K=40$, see Fig.\ref{fig1}). The results for such large $K$ system are shown in Fig.\ref{fig2} where it is observed that increasing the local connectivities by an order of magnitude does not significantly change the critical exponents of the system (see Table \ref{table1}). In fact as far as such connections remain \lq\lq local", one may expect that the critical properties of such systems remain the same when they can be regarded as \lq\lq short-range" interactions which should not effect the critical properties of the system in the thermodynamic limit. We therefore conclude that large average \textit{local} connectivity cannot by itself be responsible for mean-field behavior in neuronal avalanches since despite large local connectivities in the cortex ($K\approx 10^4$), they are a significantly small part of all possible connections of $N\approx 10^{10}$. Clearly, if one keeps increasing $K$ to levels comparable with $N$ then one expects to see mean-field exponents. But this is not a realistic limit and is not warranted by cortical samples used to study neuronal avalanches.

Small-world property of networks can also lead to mean-field behavior. We have therefore simulated the SPZ model on the Newman-Watts small-world networks. The initial networks before addition of new links are square lattices in two dimensions with $K=4$ or $K=40$. In Fig.\ref{fig3} we have plotted $s^{1.5}P(s)$ versus $s$, for different values of $q$, so that the flat horizontal portion of the plots is an indication of the mean-field behavior. Mean-field behavior expands over the entire range of data by increasing $q$, and the curves saturate for $q\geq 0.1$, i.e. they do not change by increasing $q$ above $0.1$, in both cases of $K=4$ and $K=40$. The probability distribution functions of avalanche sizes, for $0<q<0.1$, can be divided into four regions (for an indicative example see Fig.\ref{fig4}): in the first region ($s< s_{0}$) $P(s)$ does not change behavior by increasing $q$ \cite{foot1}, over the second region ($s_{0}< s< s_{b}$) regular $2D$ exponents are observed, the third region ($s_{b}< s< s_{c}$) corresponds to the mean-field behavior, and the forth region ($s\gtrsim s_{c}$) is the finite-size cutoff region. The values of $s_{0}$, $s_{b}$ and $s_{c}$ are obtained using the method explained in Fig.\ref{fig4}. As it is shown in Fig.\ref{fig5}, $s_{b}$ decreases by increasing $q$ and for $q\gtrsim 0.1$ saturates at the value of $s_{b}\approx s_{0} \approx 10^2$ where mean-field behavior dominates the entire power-law range of data ($s_{0}< s< s_{c}$) \cite{foot2}. Finite-size scaling collapses in Fig.\ref{fig6} and Fig.\ref{fig7} for size and duration of avalanches confirms the same values of mean-field exponents $\tau_{s}=1.50$ and $\tau_{d}=2.00$ for the two cases of $K=4$ and $K=40$ with $q=0.1$, see Table.\ref{table1}.

Here the important point is that regardless of the average connectivity of the system the mean-field behavior emerges, over the entire possible range of data, if at least $10$ percent of the links in the network are random long-range links. Therefore, if the mean-field exponents observed in neuronal avalanche experiments are only the result of small-world properties of the neural network in the region of interest of the brain then a considerable ratio of the links ($q=0.1$) in that region should be long-ranged. To date our information about the structural network of the brain of mammals is very incomplete \cite{sporns}, but we know that the links of the neurons in the cerebral cortex, where neuronal avalanches are measured in experiments, are mostly short ranged and the probability of connections as a function of connection range is an exponentially decaying function \cite{HHSZ}. This exponentially decaying probability of connectivities, $e^{-r/\lambda}$, between cortical neurons introduces a length scale ($\lambda$) where the likelihood of connectivity much larger than it becomes negligible. Given the above results for high local connectivity ($K=40$) along with small value of $q\lesssim 0.01$ (see Fig.\ref{fig3b}), one is led to believe that structural properties of cortical connections can not sufficiently describe the consistent and robust mean-field behavior observed in neuronal avalanches.

Another key observation against the role of long-range connections as a source of mean-field behavior in neuronal avalanches is the local, connected, wavelike spreading of instabilities seen in the experiments \cite{SACHHSCBP}. However, in our simulations with considerable amount of random connections ($q\approx 0.1$), a local instability is instantaneously transmitted via long-range links to other parts of system thus leading to multiple disconnected regions of instability, the sum of which counts as one avalanche since they have all been initiated by one instability at the seeding site.

Mean-field behavior can also be the result of noisy local dynamics in
the SPZ model. It has been shown that SPZ model on a two and three dimensional square lattice with nearest neighbor interactions exhibit
mean-field exponents for $\sigma\gtrsim 0.23$ \cite{AMIN}. In order to study the effects of the noisy dynamics on a system with high local connectivity ($K=40$),
we have plotted $s^{1.5}P(s)$ versus $s$ for different values of
$\sigma$ in Fig.\ref{fig8a}. We can see that the system approaches mean-field behavior (flat line) as $\sigma$ is increased saturating at about $\sigma\approx 0.25$, where finite-size scaling collapses of Fig.\ref{fig9} confirm the mean-field values
of $\tau_{s}\approx 1.5$, and $\tau_{d}\approx2.00$ for the critical exponents
(see Table.\ref{table1}). We note that noisy local dynamics,
regardless of the value of $K$, can result in mean-field behavior
for strong enough noise level. Therefore, this value of
$\sigma \approx 0.25$ which leads to mean-field behavior does
not depend on the connectivity ($K$) or dimension of the underlying structure \cite{AMIN}.

The amount of noise necessary to observe mean-field behavior is relatively large ($\sigma\approx 0.25$). One may think that the observed mean-field behavior in neuronal avalanches may be a result of combination of small amount of $q$ as well as large $K$ and a consequently smaller amount of noise. In order to test this we have simulated a system with $q=0.005$ and $K=40$ for various noise levels, see Fig.\ref{fig8b}. In the absence of noise ($\sigma=0.0$) one sees a two dimensional behavior for $s<s_{b}\approx 3550$ and a mean-field behavior for $s_{b}<s<s_{c}\approx 1.5\times 10^{6}$. As can be seen from the figure (and more accurately verified in Fig.\ref{fig10}) one still needs a large amount of noise ($\sigma\approx 0.25$) in order to observe mean-field behavior across the relevant range of data \cite{foot3}. This leads us to believe that noise is the key source of mean-field behavior in neuronal avalanches. This is so because as we have shown the structural properties necessary for mean-field behavior (large $K$ and $q$) are decoupled from the dynamical properties (large $\sigma$), and that such structural properties ($K\sim N$ or $q\geq 0.1$) do not seem to be validated by experimental observations.

\section{IV. CONCLUDING REMARKS}

The aim of the present work is to provide a systematic investigation into the origin of mean-field behavior observed in neuronal avalanches, occurring in resting state of various samples of the cortex, within the context of self-organized critical models. In order to do this we have modified a previously studied model of SOC and have given it neurobiological motivation and have added synaptic noise. While our conclusions are strictly true for the specific model we have studied, we believe that they are sufficiently general. However, it might be of interest to undertake similar analysis in a more realistic model of neuronal dynamics. Consequently, three leading causes have been identified and investigated: high average connectivity, random long-range connections, and synaptic noise. The first two are structural causes while the last (recently proposed) mechanism has a dynamical origin. We have modeled our neuronal dynamics based on the simple threshold dynamics of continuous variables which upon firing redistribute a random portion of their load into their predefined neighbors. This is known as the SPZ model in the SOC literature. We have also added noise into this dynamics and have studied the effect of increasing average connectivity ($K$), random long-range links ($q$), and synaptic noise ($\sigma$) on the critical behavior of this model. We find that although increasing $K$ and $q$ to arbitrary large values will lead to mean-field behavior, the typical values required to achieve this are not born out by empirical evidence in neurocortical samples. On the other hand, large enough noise will always lead to mean-field behavior regardless of the underlying structure. Large annealed noise can be well-justified specially when one considers chemical synapses which are mediated by neurotransmitters. Such processes are thought to strongly depend on many factors including the type and amount of neurotransmitters available at the time of synaptic interaction which can show a wide variability (i.e. large $\sigma$). We have therefore provided evidence for dynamical origin of mean-field behavior as opposed to conventionally accepted structural mechanisms. Our conclusion becomes more compelling when one considers that studies of cortical samples show that connection probability is of the form $e^{-r/\lambda}$ which provides a strong evidence for locality of synaptic interactions thus indicating that the structural requirements to observe mean-field behavior ($K\sim N$, or $q\gtrsim 0.1$) seem unlikely. We have therefore provided strong evidence that a dynamical origin for mean-field behavior should be seriously considered in future studies, as opposed to focusing only on structural mechanisms. In particular, avalanches in different universality classes have different avalanche exponents in $2D$, yet they yield the characteristic values $\tau_{s}=3/2$ and $\tau_{d}=2$ in mean-field theory. Hence, we might expect a similar crossover from $2D$ exponents to mean-field exponents in these other models. The details of this crossover could potentially be different in each of these models; thus, for a full understanding of this effect, it would be of interest to perform similar studies in other models of neuronal avalanches.

Furthermore, branching processes are often used in order to describe the mechanism of neuronal avalanches which, by the way, lead to mean-field exponents. We note that an avalanche can only be mapped into a branching process if each site becomes unstable only once in the process, i.e. return loops must not occur. In the SOC jargon, area and size of the avalanche must be the same, which only occur at upper critical dimension but not in two or three dimensional systems. A careful inspection of multi-electrode arrays, MEG, or fMRI images of neuronal avalanches show that in some cases a given \lq\lq site" can fire more than once in an avalanche process thus making a branching process suspect \cite{chapter}. Also, given the fact that neuronal avalanches consist of wavelike growing \lq\lq connected" regions, one can also conclude that small-world structure cannot be significant in these processes since when $q$ is appreciable avalanches break into many disconnected pieces, inconsistent with empirical observations \cite{SACHHSCBP}. The above points provide further evidence that conventional modeling of neuronal avalanches which lead to mean-field exponents are problematic and need to be reconsidered, leading further credit to motivations of the present study.

Finally, we provide some commentary on the noise in our system. Why should strong noise lead to mean-field behavior? The role of a large amplitude, annealed, zero-average noise is essentially to subtract from (or add to) the standard load of an unstable site at a given time, but due to zero average, must make that up sooner or later as the avalanche continues. This is essentially the same as subtracting a certain amount from a given site and adding that same amount to some random site at some later time. This is a random neighbor model which is well-known to provide mean-field behavior \cite{BG}. We note that this argument is independent of the particular dynamics used in the model and should be generally true. Furthermore, it is also interesting to note that one might expect that increasing the values of $K$ and $q$ (e.g. $K=40$, $q=0.005$) might have lowered the required noise level ($\sigma\approx 0.25$) in order to see mean-field exponents. This was not the case in our simulations, indicating a decoupling in dynamical and structural mechanisms for achieving mean-field behavior. We therefore believe that stochastic dynamics is the key element of mean-field behavior observed in neuronal avalanches, regardless of the actual critical neuronal dynamics (SPZ or otherwise) used, as observation of mean-field behavior has been seen in BTW model with noise \cite{Manna} and more recently in SIRS model \cite{Kinouchi}. It would be interesting to see if our results are general enough so that stochastic local dynamics will lead to mean-field behavior. And if so, it would be interesting to justify such crossover using quantitative analysis (for example exponent relations) as opposed to conceptual qualitative reasoning given above. However, we emphasize that the structural, and we believe, the dynamical mechanisms considered here are sufficient to lead to mean-field behavior once their corresponding values ($K,q,\sigma$) are large enough independent of one another, and more importantly, independent of the actual local dynamics at the sites. Therefore, it should be interesting to find out which one of these mechanisms is the dominant effect in various avalanche phenomena that exhibit mean-field behavior.

Lastly, we note that in certain studies of neuronal avalanches \cite{FIBSLD} some samples showed critical behavior, while others showed sub or supercritical  behavior. This could easily be achieved in our model by increasing (super critical) or decreasing (sub critical) the average noise from its standard zero value \cite{AMIN}. Since criticality is generally viewed as a result of intricate balance between excitation and inhibition \cite{Poil}, our zero average noise can be viewed as a mechanism which maintains such a balance while allowing for its possible violation which might be due to external as well as internal origins, as for example, imbalance of various neurotransmitters. Clearly, the inclusion of the role of inhibition in a direct way within the present model is of interest for future work and we intend to report on that elsewhere.

\begin{acknowledgments}
Support of Shiraz University Research Council is acknowledged.
\end{acknowledgments}
\bibliographystyle{apsrev}
%\bibliography{xbib}

\begin{thebibliography}{99}
\bibitem{foot1} The first two decades of data does not change behavior by increasing $q$. This indicates that for extracting real scaling behavior for large enough system sizes one can safely ignore the scaling behavior of small ($s\leq 10^{2}$) avalanches.
\bibitem{foot2} One should note that the value of $s_{b}$ is not size dependent.
\bibitem{foot3} This value of $\sigma\approx 0.25$ is unrelated to the choice of $\delta E \in [0,0.25]$ in the original model and is robust when other choices of $\delta E$ are implemented.
\end{thebibliography}

\end{document}